# Impedance Reshaping Method of DFIG System Based on Compensating Rotor Current Dynamic to Eliminate PLL Influence

Xiaoling Xiong, *Member*, *IEEE*, Bochen Luo, Longcan Li, Ziming Sun, *Member*, *IEEE*, and Frede Blaabjerg, *Fellow*, *IEEE*

***Abstract*—The phase-locked loop (PLL) used in the doubly fed induction generator (DFIG) can cause frequency coupling phenomena, which will give negative resistance characteristics of the DFIG at low frequency, resulting in stability issues under weak grid operation. Based on the multi-input-multi-output (MIMO) impedance model of DFIG system, it is found that the frequency coupling phenomena is mainly introduced by the transfer function matrix related to rotor current dynamic. This paper presents an improved impedance reshaping method based on compensating rotor current dynamic to reduce the influence of PLL, in which the rotor current dynamic is compensated before being introduced to the PI controller. Thus, the frequency coupling effect can be almost eliminated and the stability of DFIG is improved a lot. Furthermore, a simplified compensation method is proposed, which can easily be implemented. Robustness analysis is performed to illustrate the availability of the proposed methods when the system operating conditions and parameters vary. Finally, simulations based on MATLAB/Simulink are also carried out, and the results validate the effectiveness of the proposed methods.**

***Index Terms*—Doubly fed induction generator (DFIG), frequency coupling phenomena, impedance reshaping, phase-locked loop (PLL), small-signal stability.**

## I. Introduction

Wind power generation based on doubly fed induction generator (DFIG) systems has been developing rapidly [1], [2]. However, with the increasing penetration of renewable energy, the ac grid connected to them becomes weaker, giving the characteristic of a low short circuit ratio (SCR) [3]-[5]. For the DFIG system, the classic control strategy uses a phase-locked loop (PLL) to synchronize with the grid, such as orienting the point of common coupling (PCC) voltage to facilitate vector control [6], [7]. However, the asymmetric control structure of the PLL causes the frequency coupling phenomena, which may aggravate the negative resistance characteristics of the DFIG system at low frequencies, reducing the stability margin of the interconnected system [8], [9].

To solve the stability issue caused by PLL, many different methods have been proposed. One basic idea is to modify or enhance the vector control by using an extra controller. For example, a damping controller [10], [11] or a virtual impedance controller [12], [13] are proposed to suppress sub-/super-synchronous oscillations in DFIG. But the effectiveness can easily be influenced by operating conditions or system parameters. To improve the adaptability in the control system, adaptive control strategies are further proposed in [14], which can suppress the oscillation of DFIG system under different operating conditions, however, making the control structure more complex.

Other research efforts have been devoted to finding new control strategies for DFIG to replace conventional vector control based on PLL. The authors in [15], [16] adopt the direct power control (DPC) to eliminate the influence of PLL by introducing a virtual *d-q* frame rotating at a constant speed to replace the synchronous rotating *d-q* frame. However, strong frequency coupling characteristics at high frequencies are introduced by the DPC, leading to high-frequency stability issues.

In [17], a symmetrical PLL is proposed, which is able to introduce the dynamics of PLL to the *d-q* axis symmetrically, making the whole system become a single-input and single-output (SISO) system. Although the symmetrical PLL can eliminate frequency coupling phenomena and facilitate stability analysis, it still does not remove the negative resistance introduced by the PLL fundamentally. In [18], symmetrical PLL is applied to the DFIG system to eliminate the off-diagonal elements of the impedance matrix. However, the symmetrical PLL can simplify the DFIG impedance model only when other control loops are symmetrical too. Suppose the asymmetric control loops are considered, such as the dc-link voltage loop on the grid side converter (GSC). In that case, frequency coupling phenomena will still exist even if the symmetrical PLL is adopted. In [19], an impedance reshaping method based on virtual impedance is further proposed to enhance the stability of the DFIG system with symmetrical PLL.

This paper is going to put forward a general impedance reshaping method based on eliminating the dynamic of rotor current to reduce the frequency coupling phenomena of the DFIG system caused by PLL, thus improving the stability of the interconnected system. The contributions of this paper can be summarized as,

- Based on the MIMO impedance model of the DFIG system, the impact of PLL is analyzed, and their contributions to the relative transfer function matrixes are revealed. It is found that the frequency coupling phenomena is mainly caused by the transfer function matrix linked to rotor current dynamic.

- To eliminate the rotor current dynamic introduced by PLL, a second-order impedance reshaping method based on compensating rotor current dynamic is proposed, which can not only eliminate the frequency coupling phenomena caused by the PLL but also remove the negative resistance region and improve the stability margin of DFIG system.
- A first-order reshaping method is further obtained based on simplifying the second-order impedance reshaping method, which can also effectively suppress the oscillations.
- Robustness analysis is carried out, illustrating that both the proposed methods can reduce the negative resistance region and improve the stability margin of the system when the operating conditions and parameters vary.

To present all the issues, this paper is organized as follows. Section II gives DFIG system configuration, and the corresponding impedance model is presented. The contributions of different PLL-related matrices to frequency coupling phenomena are revealed. In Section III, the second-order impedance reshaping method based on compensating rotor current dynamic is proposed, and the simplified first-order method is further deduced. Section IV gives the robustness analysis to verify the effectiveness of the proposed methods when the system operating conditions and parameters vary. Section V shows the simulation results to verify the theoretical analysis. Finally, conclusions are drawn in Section VI.

## II. Impedance Model of DFIG System

### A. DFIG System Configuration

DFIG system is mainly composed of DFIG, rotor side converter (RSC), and grid side converter (GSC). The main task of DFIG+RSC is to realize maximum power point tracking and control the output power at the stator side, while the GSC is employed to stabilize the common dc-link voltage. Considering that the dc voltage is controlled to be constant by the outer-loop regulator of the GSC, the DFIG+RSC and GSC can be modeled separately, which are further paralleled to obtain the impedance model of the overall DFIG system. As the power exchange is mainly achieved by DFIG+RSC and the inductance of the filter on the GSC side is large, the amplitude of the GSC impedance will be much greater than that of DFIG+RSC. Thus, the GSC has less influence on the impedance of the DFIG system, which even can be ignored further [18]. In this paper, only the impedance of DFIG+RSC is considered.

The topology and control diagram of DFIG+RSC has shown in Fig. 1(a). $\boldsymbol{v}_{\rm sabc}$, $\boldsymbol{i}_{\rm sabc}$, $\boldsymbol{v}_{\rm rabc}$, and $\boldsymbol{i}_{\rm rabc}$ represent the voltage and current vectors at the stator and rotor side respectively, and $\boldsymbol{v}_{\rm gabc}$ is the ac grid voltage vector. The subscript abc denotes that they are in the three-phase stationary frame. The above vectors can also be expressed in a synchronous rotating frame (*d-q*), i.e., $\boldsymbol{v}_{\rm sdq}$, $\boldsymbol{i}_{\rm sdq}$, $\boldsymbol{v}_{\rm rdq}$, $\boldsymbol{i}_{\rm rdq}$, $\boldsymbol{v}_{\rm gdq}$. For example, the stator voltage vector in the two frames can be expressed as $\boldsymbol{v}_{\rm sabc} = [v_{\rm sa}\ v_{\rm sb}\ v_{\rm sc}]^{\rm T}$ and $\boldsymbol{v}_{\rm sdq} = [v_{\rm sd}\ v_{\rm sq}]^{\rm T}$. $R_{\rm g}$ and $L_{\rm g}$ are the equivalent resistance and inductance of the grid. $V_{\rm dc}$ is the dc-link voltage, which can be regarded as a constant. The rotor angular frequency $\omega_{\rm r}$ and rotor angle $\theta_{\rm r}$ are obtained by an encoder. $\theta_{\rm pll}$ is obtained by the PLL. Considering the perturbations of PCC voltage and the dynamics of PLL, there will be two different *d-q* frames, i.e., the system *d-q* frame and the control *d-q* frame [20]. Both *d-q* frames are rotating in synchronous frequency, and the relationship of them is depicted in Fig. 1(b). $\theta_{\rm s}$ is the real phase angle of the PCC voltage in the system *d-q* frame, and a phase error $\Delta\theta$ is introduced by the PLL dynamics when PCC voltage varies. To distinguish from the components in system *d-q* frame, the superscript ctrl is used to denote the components in the control *d-q* frame.

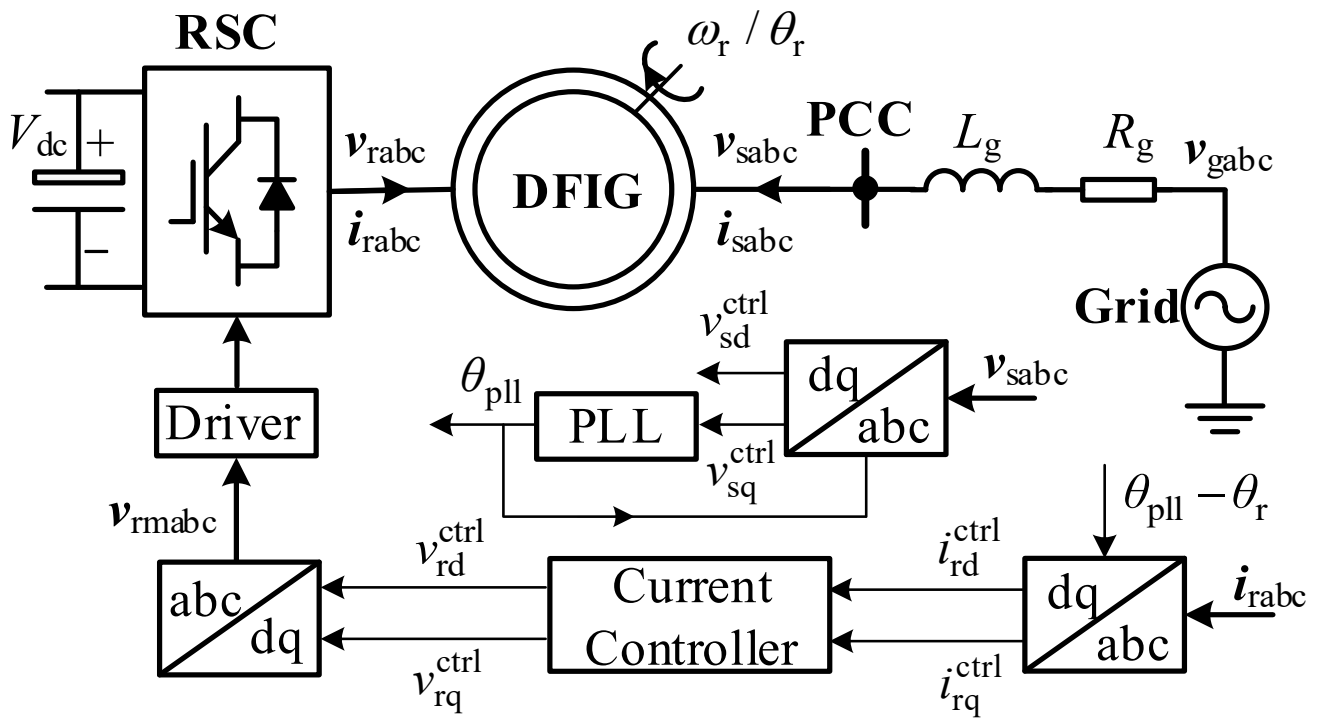

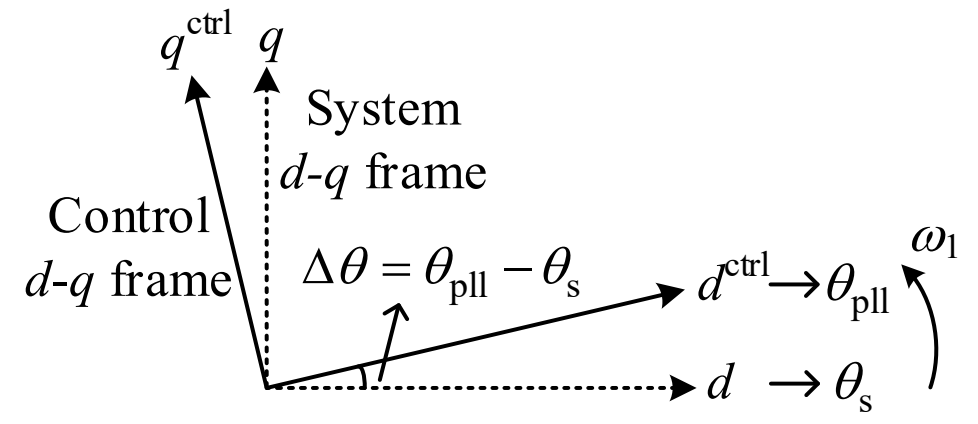

Fig. 1. The topology of DFIG system and the relationship between different *d-q* frames. (a) The topology of DFIG system connected to ac grid, (b) Diagram of the system and control *d-q* frames.

### B. Impedance Model of DFIG in d-q Frame

The voltage and flux equations of DFIG in the *d-q* frame can be expressed as follows.

$$\begin{cases} v_{\rm sd} = R_{\rm s}i_{\rm sd} + p\psi_{\rm sd} - \omega_1\psi_{\rm sq} \\ v_{\rm sq} = R_{\rm s}i_{\rm sq} + p\psi_{\rm sq} + \omega_1\psi_{\rm sd} \\ v_{\rm rd} = R_{\rm r}i_{\rm rd} + p\psi_{\rm rd} - \omega_{\rm sl}\psi_{\rm rq} \\ v_{\rm rq} = R_{\rm r}i_{\rm rq} + p\psi_{\rm rq} + \omega_{\rm sl}\psi_{\rm rd} \end{cases} \tag{1}$$

$$\begin{cases} \psi_{\rm sd} = L_{\rm s}i_{\rm sd} + L_{\rm m}i_{\rm rd} \\ \psi_{\rm sq} = L_{\rm s}i_{\rm sq} + L_{\rm m}i_{\rm rq} \\ \psi_{\rm rd} = L_{\rm m}i_{\rm sd} + L_{\rm r}i_{\rm rd} \\ \psi_{\rm rq} = L_{\rm m}i_{\rm sq} + L_{\rm r}i_{\rm rq} \end{cases} \tag{2}$$

where $p$ means differential operation. $R_{\rm s}$, $R_{\rm r}$, $L_{\rm s}$, $L_{\rm r}$, and $L_{\rm m}$ are the stator resistance, rotor resistance, stator inductance, rotor inductance, and mutual inductance, respectively. $\Psi_{\rm sd}$, $\Psi_{\rm sq}$, $\Psi_{\rm rd}$, $\Psi_{\rm rq}$, are the *d-q* components of stator flux and rotor flux. $\omega_1$ denotes the fundamental angular frequency, i.e., $100\pi$ rad/s, and $\omega_{\rm sl} = \omega_1 - \omega_{\rm r}$ is the slip angular frequency.

When the stator resistance is ignored, linearizing (1) and (2),

and applying Laplace transformation, the impedance model of DFIG in the *d-q* frame can finally be expressed as

$$\boldsymbol{Y}_{\mathrm{dq}}^{\mathrm{dfig}}=\frac{1}{L_{\mathrm{s}}}\boldsymbol{G}_{1\mathrm{dq}}+\frac{L_{\mathrm{m}}}{L_{\mathrm{s}}}\boldsymbol{G}_{1\mathrm{dq}}\boldsymbol{G}_{2\mathrm{dq}}\boldsymbol{G}_{\mathrm{pdq}} \tag{3}$$

$$\boldsymbol{G}_{1\mathrm{dq}}=\begin{bmatrix}\frac{s}{s^2+\omega_1^2} & \frac{\omega_1}{s^2+\omega_1^2}\\ \frac{-\omega_1}{s^2+\omega_1^2} & \frac{s}{s^2+\omega_1^2}\end{bmatrix};\ \boldsymbol{G}_{2\mathrm{dq}}=\frac{L_{\mathrm{m}}}{L_{\mathrm{s}}}\begin{bmatrix}s & -\omega_{\mathrm{sl}}\\ \omega_{\mathrm{sl}} & s\end{bmatrix};$$

$$\boldsymbol{G}_{\mathrm{pdq}}=\begin{bmatrix}\frac{(R_{\mathrm{r}}+sL_{\mathrm{r}}\sigma)}{(R_{\mathrm{r}}+sL_{\mathrm{r}}\sigma)^2+(\omega_{\mathrm{sl}}L_{\mathrm{r}}\sigma)^2} & \frac{\omega_{\mathrm{sl}}L_{\mathrm{r}}\sigma}{(R_{\mathrm{r}}+sL_{\mathrm{r}}\sigma)^2+(\omega_{\mathrm{sl}}L_{\mathrm{r}}\sigma)^2}\\ \frac{-\omega_{\mathrm{sl}}L_{\mathrm{r}}\sigma}{(R_{\mathrm{r}}+sL_{\mathrm{r}}\sigma)^2+(\omega_{\mathrm{sl}}L_{\mathrm{r}}\sigma)^2} & \frac{(R_{\mathrm{r}}+sL_{\mathrm{r}}\sigma)}{(R_{\mathrm{r}}+sL_{\mathrm{r}}\sigma)^2+(\omega_{\mathrm{sl}}L_{\mathrm{r}}\sigma)^2}\end{bmatrix} \tag{4}$$

where $s$ denote the Laplace operator and $\sigma = 1 - L_{\mathrm{m}}^2/(L_{\mathrm{s}}L_{\mathrm{r}})$.

It can be found that the diagonal elements of $\boldsymbol{G}_{1\mathrm{dq}}$, $\boldsymbol{G}_{2\mathrm{dq}}$, and $\boldsymbol{G}_{\mathrm{pdq}}$ are equal, and their off-diagonal elements are opposite, indicating that the DFIG is a symmetrical system without PLL and RSC, which can also be considered as a SISO system using complex vectors model.

## *C. Impedance Model of DFIG System with PLL*

The influence of PLL on stator voltage, rotor current, and rotor voltage can be derived as [9]

$$\begin{aligned}\boldsymbol{v}_{\mathrm{sdq}}^{\mathrm{ctrl}}&=\boldsymbol{v}_{\mathrm{sdq}}-\boldsymbol{G}_{\mathrm{PLL}}^{\mathrm{v}}\cdot\boldsymbol{v}_{\mathrm{sdq}}\\ \boldsymbol{i}_{\mathrm{rdq}}^{\mathrm{ctrl}}&=\boldsymbol{i}_{\mathrm{rdq}}-\boldsymbol{G}_{\mathrm{PLL}}^{\mathrm{i}}\cdot\boldsymbol{v}_{\mathrm{sdq}}\\ \boldsymbol{v}_{\mathrm{rdq}}&=\boldsymbol{v}_{\mathrm{rdq}}^{\mathrm{ctrl}}+\boldsymbol{G}_{\mathrm{PLL}}^{\mathrm{m}}\cdot\boldsymbol{v}_{\mathrm{sdq}}\end{aligned} \tag{5}$$

where the *d-q* components all represent the small signal perturbation in the *s*-domain, and the matrixes related to the PLL dynamics can be expressed as

$$\begin{aligned}\boldsymbol{G}_{\mathrm{PLL}}^{\mathrm{v}}&=\begin{bmatrix}0 & -V_{\mathrm{sq0}}H_{\mathrm{pll}}(s)\\ 0 & V_{\mathrm{sd0}}H_{\mathrm{pll}}(s)\end{bmatrix}\\ \boldsymbol{G}_{\mathrm{PLL}}^{\mathrm{i}}&=\begin{bmatrix}0 & -I_{\mathrm{rq0}}H_{\mathrm{pll}}(s)\\ 0 & I_{\mathrm{rd0}}H_{\mathrm{pll}}(s)\end{bmatrix}\\ \boldsymbol{G}_{\mathrm{PLL}}^{\mathrm{m}}&=\begin{bmatrix}0 & -V_{\mathrm{rq0}}H_{\mathrm{pll}}(s)\\ 0 & V_{\mathrm{rd0}}H_{\mathrm{pll}}(s)\end{bmatrix}\end{aligned} \tag{6}$$

Here, $V_{\mathrm{sd0}}$, $V_{\mathrm{sq0}}$, $I_{\mathrm{rd0}}$, $I_{\mathrm{rq0}}$, $V_{\mathrm{rd0}}$, $V_{\mathrm{rq0}}$ denote the steady-state values. $H_{\mathrm{pll}}(s)$ is a second-order transfer function for the PLL, which can be expressed as [21]

$$H_{\mathrm{pll}}(s)=\frac{K_{\mathrm{pPLL}}s+K_{\mathrm{iPLL}}}{s^2+K_{\mathrm{pPLL}}V_{\mathrm{sd0}}s+K_{\mathrm{iPLL}}V_{\mathrm{sd0}}} \tag{7}$$

where $K_{\mathrm{pPLL}}$ and $K_{\mathrm{iPLL}}$ are the proportional gain and integral gain of the PLL controller.

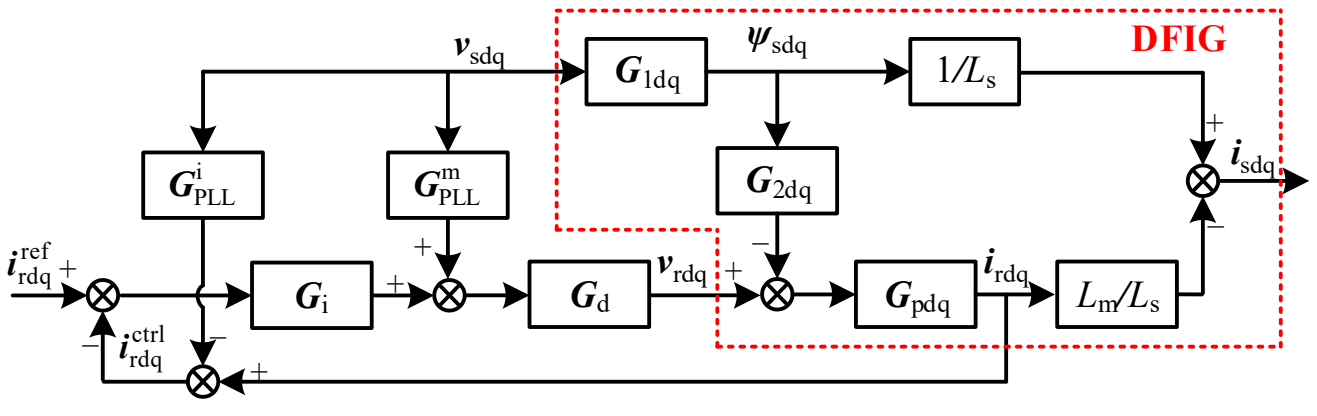

Fig. 2. Impedance Model of DFIG system in d-q frame with considering PLL.

Considering the dynamics of PLL, and the rotor current control loop of RSC, the impedance model of the DFIG system can be depicted in Fig. 2.

In Fig. 2, $\boldsymbol{G}_{\mathrm{i}}$ and $\boldsymbol{G}_{\mathrm{d}}$ present the transfer matrixes for PI controller and system delay in the RSC current control loop. According to Fig. 2, the admittance of DFIG system in the *d-q* frame can be deduced as

$$\begin{aligned}\boldsymbol{Y}_{\mathrm{dq}}^{\mathrm{dfrc}}=\frac{\boldsymbol{i}_{\mathrm{sdq}}}{\boldsymbol{v}_{\mathrm{sdq}}}=&\underbrace{\frac{1}{L_{\mathrm{s}}}\boldsymbol{G}_{1\mathrm{dq}}+\frac{L_{\mathrm{m}}}{L_{\mathrm{s}}}\left(\boldsymbol{I}+\boldsymbol{G}_{\mathrm{pdq}}\boldsymbol{G}_{\mathrm{d}}\boldsymbol{G}_{\mathrm{i}}\right)^{-1}\boldsymbol{G}_{\mathrm{pdq}}\boldsymbol{G}_{2\mathrm{dq}}\boldsymbol{G}_{1\mathrm{dq}}}_{\boldsymbol{Y}_{\mathrm{dq}}^{\mathrm{SISO}}}\\ &\underbrace{-\frac{L_{\mathrm{m}}}{L_{\mathrm{s}}}\left(\boldsymbol{I}+\boldsymbol{G}_{\mathrm{pdq}}\boldsymbol{G}_{\mathrm{d}}\boldsymbol{G}_{\mathrm{i}}\right)^{-1}\boldsymbol{G}_{\mathrm{pdq}}\boldsymbol{G}_{\mathrm{d}}\boldsymbol{G}_{\mathrm{i}}\boldsymbol{G}_{\mathrm{PLL}}^{\mathrm{i}}}_{\boldsymbol{Y}_{\mathrm{dq}}^{\mathrm{i}}}\underbrace{-\frac{L_{\mathrm{m}}}{L_{\mathrm{s}}}\left(\boldsymbol{I}+\boldsymbol{G}_{\mathrm{pdq}}\boldsymbol{G}_{\mathrm{d}}\boldsymbol{G}_{\mathrm{i}}\right)^{-1}\boldsymbol{G}_{\mathrm{pdq}}\boldsymbol{G}_{\mathrm{d}}\boldsymbol{G}_{\mathrm{PLL}}^{\mathrm{m}}}_{\boldsymbol{Y}_{\mathrm{dq}}^{\mathrm{m}}}\end{aligned} \tag{8}$$

From (8), it can be seen that $\boldsymbol{Y}_{\mathrm{dq}}^{\mathrm{dfrc}}$ consists of three parts, i.e., $\boldsymbol{Y}_{\mathrm{dq}}^{\mathrm{SISO}}$, $\boldsymbol{Y}_{\mathrm{dq}}^{\mathrm{i}}$, and $\boldsymbol{Y}_{\mathrm{dq}}^{\mathrm{m}}$. $\boldsymbol{Y}_{\mathrm{dq}}^{\mathrm{SISO}}$ is symmetrical and can be transferred to a SISO impedance using complex vectors model. The effects of PLL on rotor current and rotor voltage are included in $\boldsymbol{Y}_{\mathrm{dq}}^{\mathrm{i}}$ and $\boldsymbol{Y}_{\mathrm{dq}}^{\mathrm{m}}$. Therefore, the contributions of $\boldsymbol{Y}_{\mathrm{dq}}^{\mathrm{i}}$ and $\boldsymbol{Y}_{\mathrm{dq}}^{\mathrm{m}}$ to the frequency coupling phenomena can be studied separately.

On this basis, the admittance of the DFIG system in *d-q* frame can be further converted in the stationary domain as [22]

$$\boldsymbol{Y}_{\alpha\beta}^{\mathrm{dfrc}}(s)=\frac{1}{2}\begin{bmatrix}1 & j\\ 1 & -j\end{bmatrix}\boldsymbol{Y}_{\mathrm{dq}}^{\mathrm{dfrc}}(s-j\omega_1)\begin{bmatrix}1 & 1\\ -j & j\end{bmatrix} \tag{9}$$

Similarly, the admittances $\boldsymbol{Y}_{\alpha\beta}^{\mathrm{SISO}}$, $\boldsymbol{Y}_{\alpha\beta}^{\mathrm{i}}$, and $\boldsymbol{Y}_{\alpha\beta}^{\mathrm{m}}$ in the stationary domain can also be obtained, and the relationship between them can be expressed as

$$\boldsymbol{Y}_{\alpha\beta}^{\mathrm{dfrc}}=\boldsymbol{Y}_{\alpha\beta}^{\mathrm{SISO}}+\boldsymbol{Y}_{\alpha\beta}^{\mathrm{i}}+\boldsymbol{Y}_{\alpha\beta}^{\mathrm{m}} \tag{10}$$

## *D. Verification and Analysis of the Impedance Model*

TABLE I
PARAMETERS OF DFIG SYSTEM USED IN SIMULATION

| Symbol | Parameter | Value |
|---|---|---|
| $U_{\mathrm{N}}$ | Rated voltage | 1 p.u.( 690 V) |
| $P_{\mathrm{N}}$ | Rated power | 1 p.u. (1.5 MW) |
| $f_1$ | Fundamental frequency | 1 p.u. (50 Hz) |
| $f_{\mathrm{r}}$ | Rotor frequency | 1.1 p.u. (55 Hz) |
| $n_{\mathrm{p}}$ | Pole pairs | 2 |
| $V_{\mathrm{dc}}$ | DC-link voltage | 1 p.u. (1150 V) |
| $L_{\mathrm{ls}}$ | Stator leakage | 0.059 p.u. |
| $L_{\mathrm{lr}}$ | Rotor leakage | 0.082 p.u. |
| $L_{\mathrm{ms}}$ | Mutual inductance | 2.919 p.u. |
| $R_{\mathrm{s}}$ | Stator resistance | 0.0076 p.u. |
| $R_{\mathrm{r}}$ | Rotor resistance | 0.0063 p.u. |
| $K_{\mathrm{e}}$ | Turns ratio | 0.33 |
| $T_{\mathrm{s}}$ | Switching period | 0.1 ms |
| $K_{\mathrm{pPLL}}$ | Proportional gain of PLL controller | 0.48 p.u. |
| $K_{\mathrm{iPLL}}$ | Integral gain of PLL controller | 20.85 p.u. |
| $K_{\mathrm{pI}}$ | Proportional gain of current controller | 15.75 p.u. |
| $K_{\mathrm{iI}}$ | Integral gain of current controller | 1575.3 p.u. |

The simulation model of a DFIG system with PLL under a weak grid is developed in MATLAB/Simulink, and the detailed parameters are shown in Table I. As shown in Fig. 3, the analytical model of the DFIG system, i.e., $\boldsymbol{Y}_{\alpha\beta}^{\mathrm{dfrc}}$, matches the simulation results well, which validates the accuracy of the DFIG impedance model given in (9) and (10).

Moreover, the analytical models for different parts combinations are also given in Fig. 3. From there, it can be seen that the Bode diagram of $\boldsymbol{Y}_{\alpha\beta}^{\mathrm{SISO}} + \boldsymbol{Y}_{\alpha\beta}^{\mathrm{i}}$ is almost coincident with $\boldsymbol{Y}_{\alpha\beta}^{\mathrm{dfrc}}$, indicating $\boldsymbol{Y}_{\alpha\beta}^{\mathrm{m}}$ has little influence on the MIMO admittance of the DFIG system. Nevertheless, the diagonal elements of $\boldsymbol{Y}_{\alpha\beta}^{\mathrm{SISO}} + \boldsymbol{Y}_{\alpha\beta}^{\mathrm{m}}$ are coincident with $\boldsymbol{Y}_{\alpha\beta}^{\mathrm{SISO}}$ and the off-diagonal elements of it decrease to below −20 dB, which is much lower than the diagonal elements and it can almost be neglected. Therefore, the frequency coupling phenomena caused by the PLL is mainly related to $\boldsymbol{Y}_{\alpha\beta}^{\mathrm{i}}$. When $\boldsymbol{Y}_{\alpha\beta}^{\mathrm{i}}$ is ignored, the influence of the PLL will be greatly reduced, and the whole DFIG system can be approximately regarded as a SISO system using complex vectors model.

Basically, the rated power of DFIG is about or beyond 1.5 MW, and its rated voltage is about 690 V (960 V for higher power DFIG), which leads to the rated current being much greater than the rated voltage, especially on the rotor side. Thus, the steady-state values in $\boldsymbol{G}_{\mathrm{PLL}}^{\mathrm{i}}$ will be much greater than that in $\boldsymbol{G}_{\mathrm{PLL}}^{\mathrm{m}}$, finally resulting in $\boldsymbol{Y}_{\alpha\beta}^{\mathrm{i}}$ contributing more than $\boldsymbol{Y}_{\alpha\beta}^{\mathrm{m}}$ to frequency coupling phenomena.

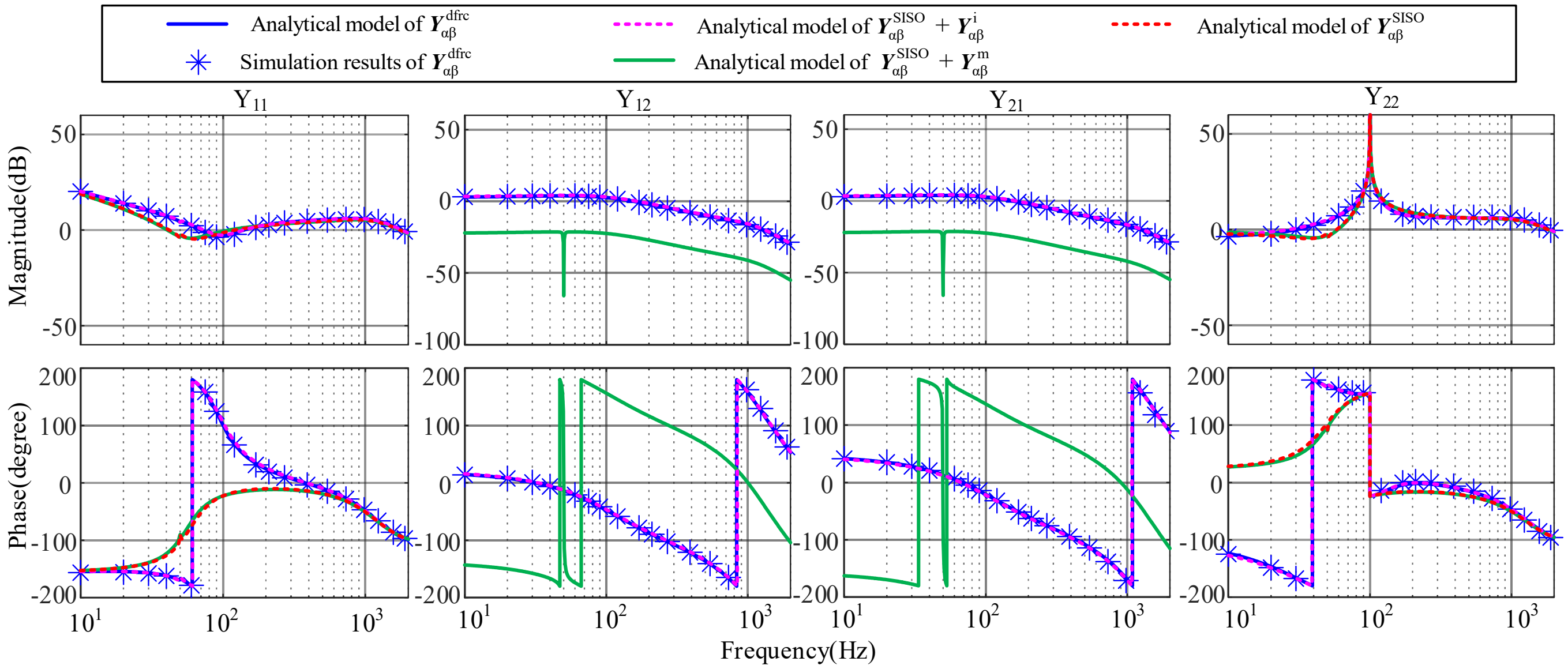

Fig. 3. Comparison of the DFIG impedance models with different parts combinations.

## III. Impedance Reshaping Method

### *A. Proposal of Impedance Reshaping Method*

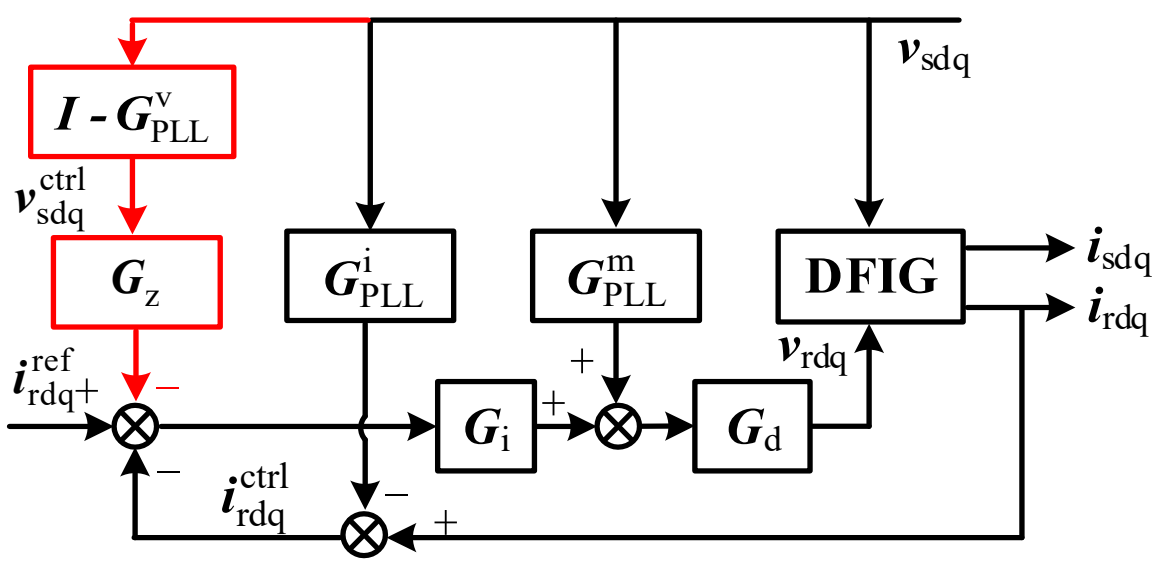

Fig. 4. Control diagram of the DFIG system with reshaping to reduce frequency coupling phenomenon.

The abovementioned analysis shows that the frequency coupling phenomena is mainly caused by $\boldsymbol{Y}_{\alpha\beta}^{\mathrm{i}}$, which is dominated by the matrix $\boldsymbol{G}_{\mathrm{PLL}}^{\mathrm{i}}$ that is related to rotor current dynamic. Therefore, this paper tries to use a parallel impedance reshaping path to compensate for rotor current dynamic caused by PLL, thus eliminating the influence of $\boldsymbol{G}_{\mathrm{PLL}}^{\mathrm{i}}$, as shown in Fig. 4. It can be seen that the introduced path includes two blocks, i.e., $\boldsymbol{I}$-$\boldsymbol{G}_{\mathrm{PLL}}^{\mathrm{v}}$ and $\boldsymbol{G}_{\mathrm{z}}$. As $\boldsymbol{I}$-$\boldsymbol{G}_{\mathrm{PLL}}^{\mathrm{v}}$ is necessary for transforming the stator voltage to the control *d-q* frame, thus, only $\boldsymbol{G}_{\mathrm{z}}$ should be designed.

In order to eliminate the influence of $\boldsymbol{G}_{\mathrm{PLL}}^{\mathrm{i}}$, the following relationship needs to be satisfied

$$\boldsymbol{G}_{\mathrm{z}} = \boldsymbol{G}_{\mathrm{PLL}}^{\mathrm{i}}\left(\boldsymbol{I} - \boldsymbol{G}_{\mathrm{PLL}}^{\mathrm{v}}\right)^{-1} = \begin{bmatrix} 0 & -I_{\mathrm{rq0}}\left(K_{\mathrm{pPLL}}s + K_{\mathrm{iPLL}}\right)/s^2 \\ 0 & I_{\mathrm{rd0}}\left(K_{\mathrm{pPLL}}s + K_{\mathrm{iPLL}}\right)/s^2 \end{bmatrix} \quad (11)$$

where $\boldsymbol{I}$ represents a 2×2 identity matrix.

From (11), it can be seen that $\boldsymbol{G}_{\mathrm{z}}$ has an asymmetric structure, and its non-zero elements exist only in the second column, which is related to $v_{\mathrm{sq}}^{\mathrm{ctrl}}$. Besides, $\boldsymbol{G}_{\mathrm{z}}$ is similar to a second-order integral term, and the dc components of $v_{\mathrm{sq}}^{\mathrm{ctrl}}$ will be amplified, causing the rotor current to deviate from its expected values. Thus, it is necessary to add a second-order high pass filter (HPF) in the reshaping branch to remove the dc components of $v_{\mathrm{sq}}^{\mathrm{ctrl}}$. At the same time, to improve the adaptability of the proposed method, the steady-state values can be replaced by reference values, i.e., $\boldsymbol{I}_{\mathrm{rdqref}}$. As a result, $\boldsymbol{G}_{\mathrm{z}}$ can be deduced as

$$\boldsymbol{G}_{\mathrm{z1}} = G_{\mathrm{HPF}}(s)\begin{bmatrix} 0 & -I_{\mathrm{rqref}}\left(K_{\mathrm{pPLL}}s + K_{\mathrm{iPLL}}\right)/s^2 \\ 0 & I_{\mathrm{rdref}}\left(K_{\mathrm{pPLL}}s + K_{\mathrm{iPLL}}\right)/s^2 \end{bmatrix} \quad (12)$$

where $G_{\mathrm{HPF}}(s)$ represents the transfer function of the second-order HPF, which can be written as

$$G_{\text{HPF}}(s) = \frac{A(\infty)s^2}{s^2 + (\omega_{\text{HPF}}/Q)s + \omega_{\text{HPF}}^2} \tag{13}$$

To ensure that the HPF will not greatly affect the decoupling function of the proposed method, the cut-off frequency $\omega_{\text{HPF}}$ should be taken as small as possible, which is set to $2\pi \cdot 1$ rad/s in this paper. Accordingly, the gain coefficient $A(\infty)$ and the quality factor $Q$ are both set as 1.

For the DFIG system, most of the oscillations usually occur beyond 50 Hz [23]. Thus, the high-frequency range, i.e., above 50 Hz, is mainly in focus here. In this frequency range, if in some cases $K_{\text{pPLL}}s$ is much greater than $K_{\text{iPLL}}$, the integral gain $K_{\text{iPLL}}$ can thus be ignored for simplification. To this end, the reshaping method in (12) can be further simplified as

$$\boldsymbol{G}_{z2} = \frac{s}{s+\omega_{\text{HPF}}}\begin{bmatrix} 0 & \dfrac{-I_{\text{rqref}}K_{\text{pPLL}}}{s} \\ 0 & \dfrac{I_{\text{rdref}}K_{\text{pPLL}}}{s}\end{bmatrix} = \begin{bmatrix} 0 & \dfrac{-I_{\text{rqref}}K_{\text{pPLL}}}{s+\omega_{\text{HPF}}} \\ 0 & \dfrac{I_{\text{rdref}}K_{\text{pPLL}}}{s+\omega_{\text{HPF}}}\end{bmatrix} \tag{14}$$

Finally, the original second-order reshaping method becomes a first-order one, and only a first-order HPF is required. By further derivation, just a pair of asymmetrical low-pass filters (LPF) should be added, which greatly decreases the implementation complexity. The implementation diagram is shown in Fig. 5, where the reshaping branch is plotted in red. $G_z^{12}(s)$ and $G_z^{22}(s)$ represent the second column elements of $\boldsymbol{G}_{z1}$ or $\boldsymbol{G}_{z2}$. $K_{\text{pI}}$ and $K_{\text{iI}}$ denote the proportional gain and integral gain of the current controller.

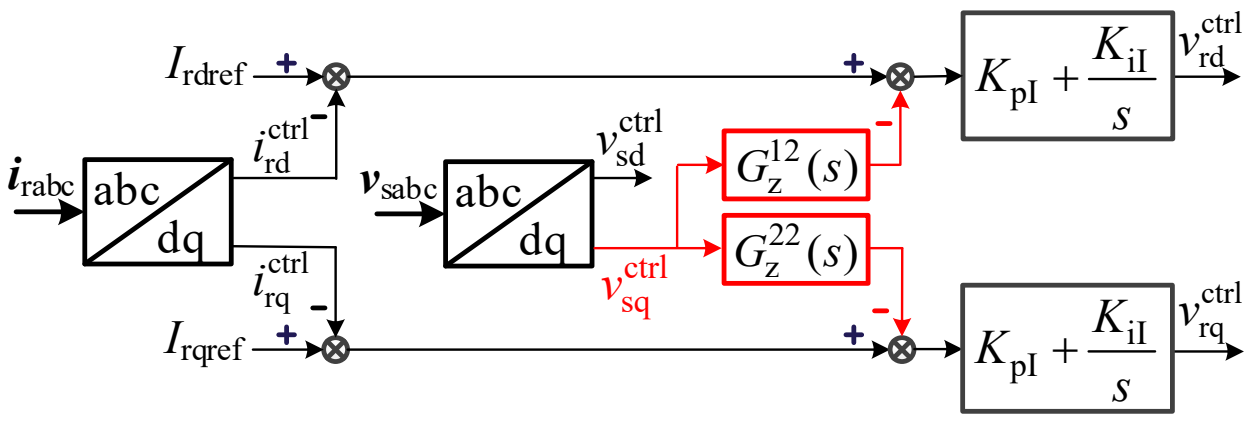

Fig. 5. The implementation diagram of RSC current controller with adopting the proposed methods.

It should be noted that the compensation methods for rotor current dynamic in this paper are proposed to cancel frequency coupling phenomena and reshape impedance model, resulting in the stability improvement. This is different from the existing methods based on the voltage feedforward or adding particular filters in the voltage feedforward branches [24], [25], where the feedforward terms are usually behind the current PI controller and used to optimize impedance characteristics or improve the response of tracking reference signal. In this paper, the reshaping terms are added before PI controller to compensate for the dynamic of the rotor current, thus reducing the influence of PLL. Meanwhile, the proposed methods have design-oriented parameters.

## B. Stability Analysis

The stability of the DFIG interconnected system can be assessed through the generalized Nyquist criterion (GNC) [26], i.e., analyzing the eigen-loci, which can be calculated as

$$\det(\lambda \boldsymbol{I} - \boldsymbol{Y}_{\alpha\beta}^{\text{dfrc}} \boldsymbol{Z}_{\text{g}}) \tag{15}$$

where $\boldsymbol{Z}_{\text{g}}$ represents the grid impedance matrix [27].

Assuming the DFIG interconnected system operates at the rated power, as shown in Table I. Fig. 6 shows the generalized Nyquist plots of the system when the proportional gain of PLL increases from 1 p.u. to 2.57 p.u. with SCR = 2.

From Fig. 6(a), it can be seen that when $K_{\text{pPLL}}$ is 1 p.u., the eigen-loci of the system do not encircle the critical point (−1, $j$0), representing that the system is stable. However, when $K_{\text{pPLL}}$ increases to 2.57 p.u., (−1, $j$0) point is encircled, indicating an oscillation will occur. From the enlarged 3D-view Nyquist diagram in Fig. 6(b), the oscillation frequencies are about 114 Hz and −14 Hz, which is consistent with the frequency coupling phenomena mentioned in [22].

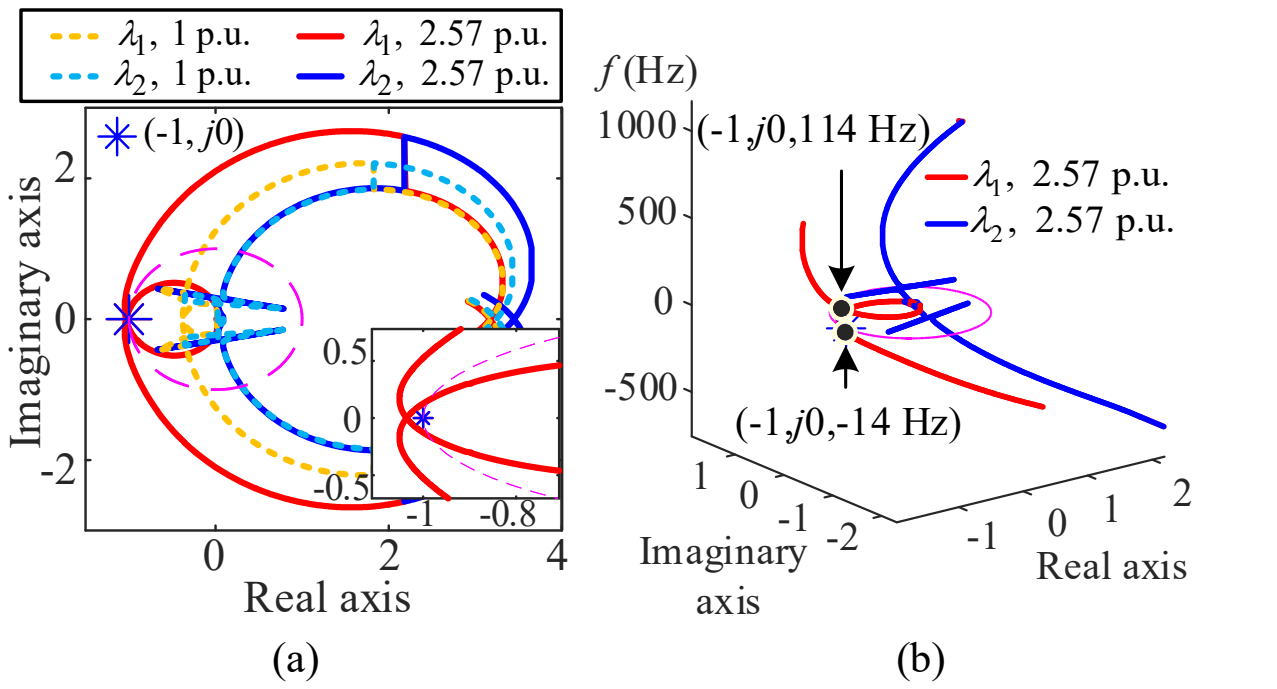

Fig. 6. Nyquist diagram of DFIG interconnected system. (a) $K_{\text{pPLL}}$ increases from 1 p.u. to 2.57 p.u., (b) The enlarged 3D-view for $K_{\text{pPLL}}$ = 2.57 p.u..

In order to acquire the stability margin of the system directly, the equivalent SISO impedance model [28] of the interconnected system should be deduced, and the positive-sequence one is calculated as

$$Z_{\text{peq}} = Z_{11} - \frac{Z_{21}Z_{12}}{Z_{22} + Z_{\text{g22}}}$$
$$Z_{\text{pgeq}} = Z_{\text{g11}} \tag{16}$$

where $Z_{11}$, $Z_{12}$, $Z_{21}$, $Z_{22}$ are the elements of $(\boldsymbol{Y}_{\alpha\beta}^{\text{dfrc}})^{-1}$, respectively. The subscript numbers represent the corresponding row and column position. $Z_{\text{g11}}$ and $Z_{\text{g22}}$ are the diagonal elements of $\boldsymbol{Z}_{\text{g}}$. From (16), it can be clearly seen that the system will be a real SISO system, i.e., $Z_{\text{peq}} = Z_{11}$, when the frequency coupling effect is eliminated. In other words, $Z_{12} \approx 0$ and $Z_{21} \approx 0$. Similarly, the equivalent negative-sequence SISO impedance model can also be derived using the same method, which is ignored for simplification.

The Bode diagram of the equivalent SISO impedance model with or without adopting the proposed methods is given in Fig. 7. When $K_{\text{pPLL}}$ is 2.57 p.u., from Fig. 7(a), the amplitude-frequency characteristic curves of $Z_{\text{peq}}$ and $Z_{\text{pgeq}}$ intersect at 114 Hz, and the corresponding phase difference is 180.9°, indicating an oscillation will happen at positive 114 Hz. Meanwhile, an oscillation at negative 14 Hz will also happen, concluded by the negative-sequence impedance model. The results also agree well with the analysis using the MIMO impedance model in Fig. 6. Fortunately, when the reshaping method $\boldsymbol{G}_{z1}$ is adopted, the Bode diagram of $Z_{\text{peq}}$ is basically coincident with the curves of $Z_{11}$, indicating that the frequency coupling effect was almost eliminated. Meanwhile, the negative resistance region can be removed, and the stability of the system is improved a lot, as seen in Fig. 7(a). The intersection frequency increases to 227 Hz, and the corresponding phase difference decreases to 75.8°.

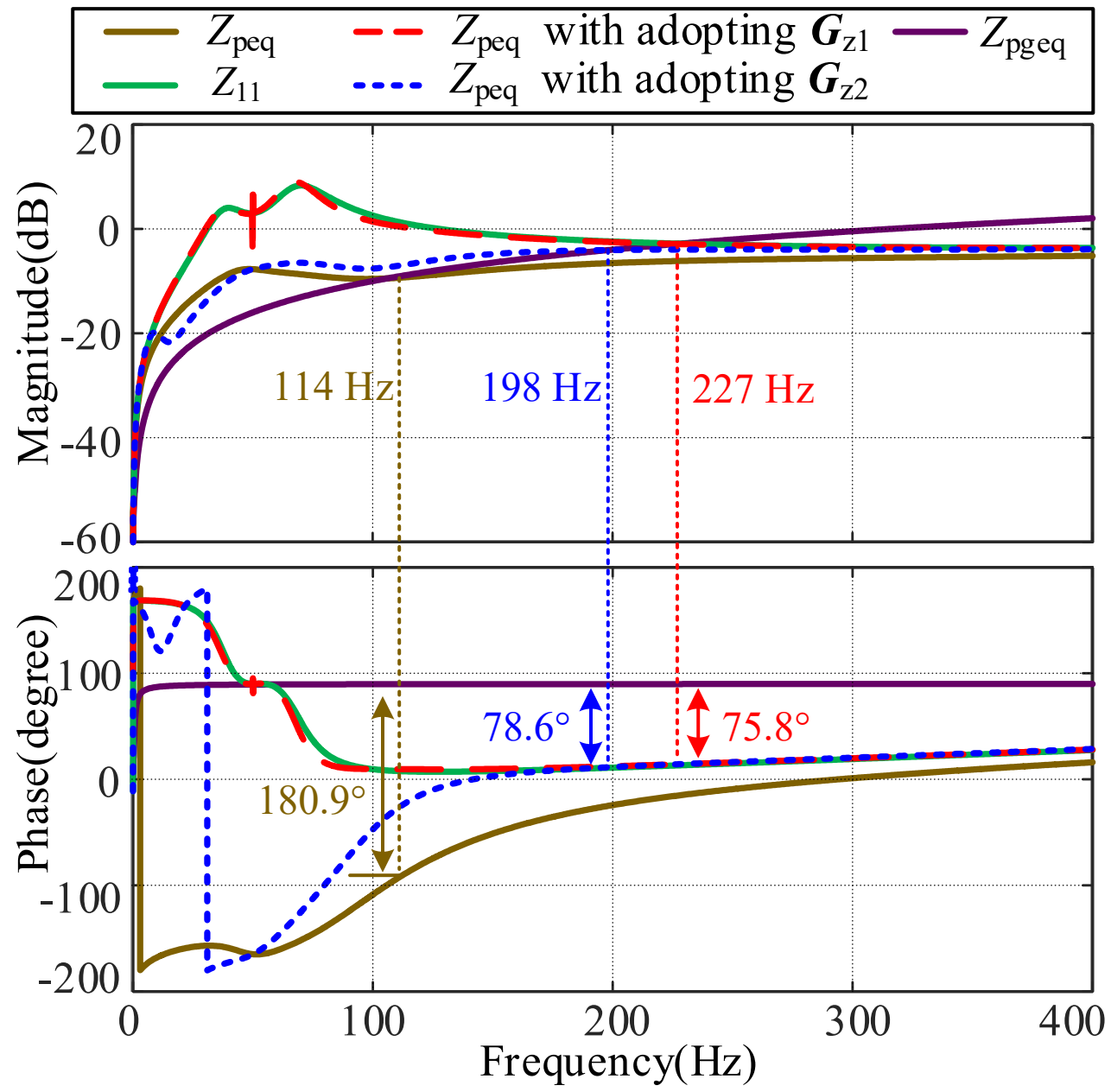

(a)

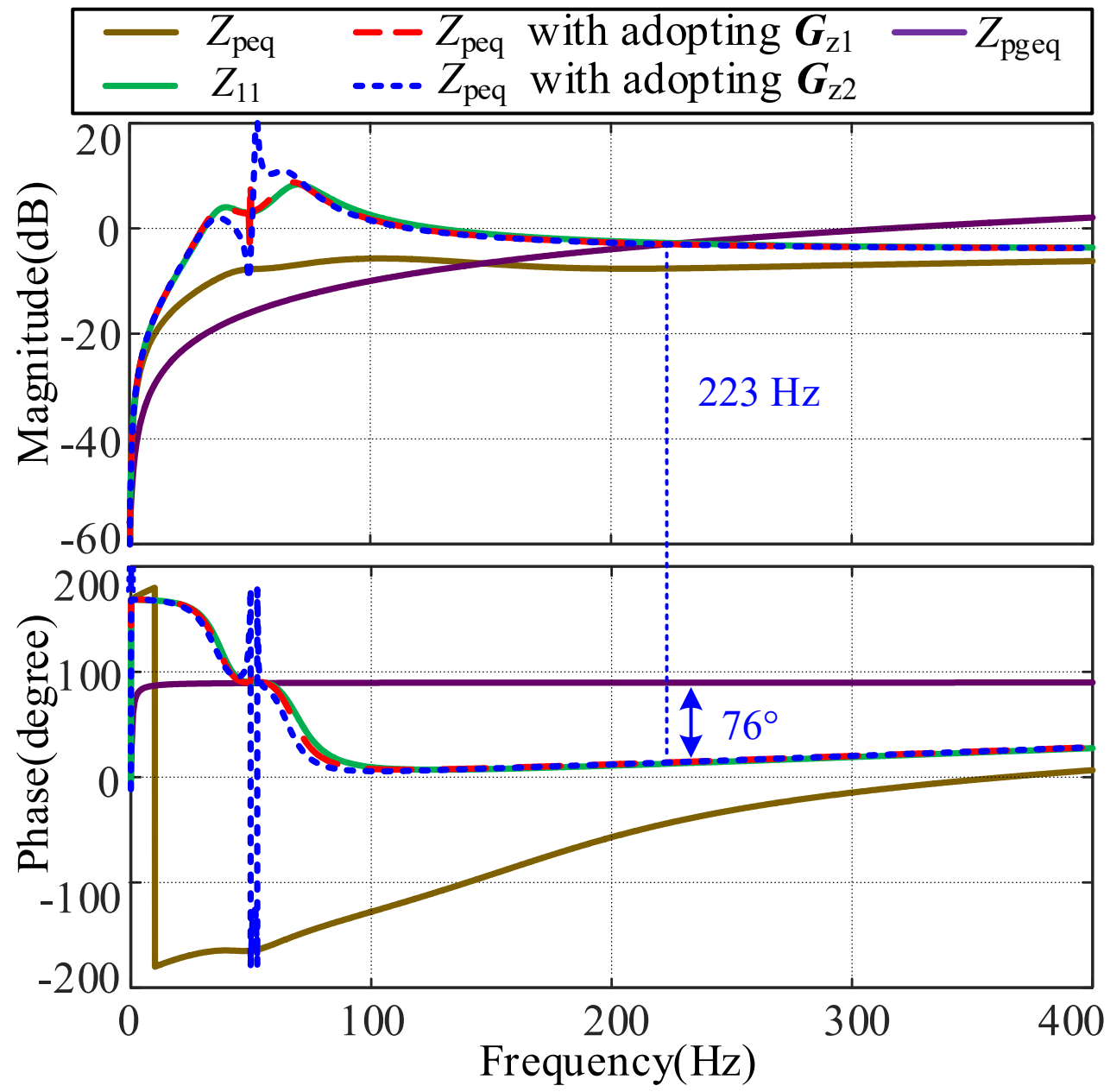

(b)

Fig. 7. Bode diagram of the equivalent SISO impedance model with adopting proposed reshaping methods. (a) $K_{pPLL}$ = 2.57 p.u., (b) $K_{pPLL}$ = 5 p.u..

When the simplified method $\boldsymbol{G}_{z2}$ is adopted, from Fig. 7(a), the negative resistance region of $Z_{peq}$ will be also reduced. Therefore, the phase difference decreases to 78.6° at the intersection frequency 198 Hz, resulting in the stability margin of the system increased a lot. If $K_{pPLL}$ is increased to 5 p.u., i.e., $K_{pPLL}s >> K_{iPLL}$ is almost satisfied, the Bode diagram of $Z_{peq}$ with adopting $\boldsymbol{G}_{z2}$ will also be coincident with $Z_{11}$, as shown in Fig. 7(b). From here, it can be seen that the negative resistance region will be removed, showing $\boldsymbol{G}_{z2}$ plays the same role as $\boldsymbol{G}_{z1}$ under this condition.

The above analysis indicates that $\boldsymbol{G}_{z1}$ and $\boldsymbol{G}_{z2}$ can both improve the stability of the DFIG system. Meanwhile, $\boldsymbol{G}_{z1}$ can almost remove all the negative resistance regions under any conditions, while $\boldsymbol{G}_{z2}$ can only achieve this when $K_{pPLL}$ is relatively large.

## IV. Robust Analysis

When the DFIG system is connected to a weak grid, the system parameters may vary, such as SCR, active and reactive power requirements, rotor frequency, DFIG leakage inductance, and so on. Therefore, it is necessary to analyze the robustness of the proposed methods against the system parameters or operating condition variations.

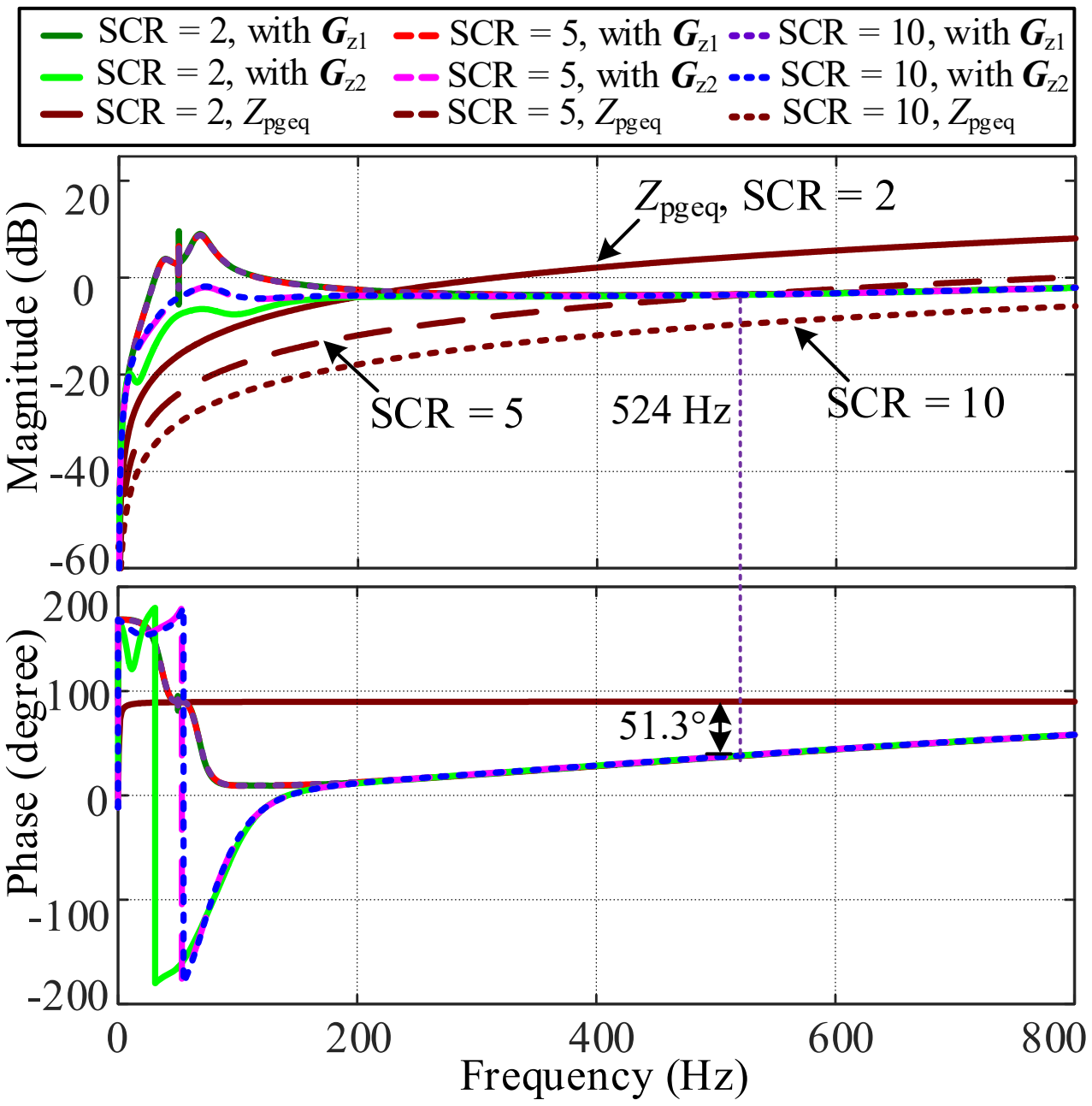

Fig. 8. Bode diagram of equivalent SISO impedance model of the DFIG system adopting the reshaping methods with different SCR.

When $\boldsymbol{G}_{z1}$ or $\boldsymbol{G}_{z2}$ is employed, the Bode diagrams of $Z_{peq}$ with different SCR are given in Fig. 8. As there is no equilibrium point for the steady-state when SCR < 2, SCR = 2 is thus regarded as the worst grid condition. Since the methods are proven to be effective with SCR = 2, it is only necessary to study the effectiveness of the methods when SCR increases. From Fig. 8, when the SCR increases to 5, the amplitude-frequency curves of $Z_{peq}$ and $Z_{pgeq}$ will intersect at about 524 Hz, and the phase difference is 51.3°, denoting the system is quite stable. Meanwhile, the intersection frequency of $Z_{peq}$ and $Z_{pgeq}$ will be much higher when the SCR increases to 10, where the DFIG system is in a positive resistance region, and the system has no risk of oscillation. The above analysis implies that as long as the parameters of the impedance reshaping methods are designed for the lowest SCR, the system can still keep stable when the SCR varies.

The robust analysis against other parameters deviation, such as power requirements, rotor frequency, and DFIG leakage inductance, is also investigated, by plotting and analyzing the Bode diagram of $Z_{peq}$ and $Z_{pgeq}$, just like the process in Fig. 8. The corresponding results are shown in Table II.

Generally, the power factor of the DFIG is kept between 0.95 to 1. However, when the voltage sag occurs, additional reactive power is required by DFIG to support the PCC voltage. As it can be seen from Table II, when DFIG operates at the power factor of 0.95 or generates reactive power at −1 p.u., the phase differences are both smaller than 180° at the intersection frequency, meaning both methods still do well when the power requirements vary in a prospective range. Similarly, a different rotor frequency range (deviating ±20%) and DFIG leakage inductance deviating ±20% from the rated value are analyzed, and the phase differences are also calculated for the intersection frequency. It can be found that the phase differences are still far less than 180°, implying that $\boldsymbol{G}_{z1}$ and $\boldsymbol{G}_{z2}$ are both effective for rotor frequency and leakage inductance varying in a normal range.

TABLE II
ROBUSTNESS ANALYSIS RESULTS AGAINST PARAMETERS DEVIATION

| Robustness analysis against power reference deviations | | |
|---|---|---|
| Method | Power requirements | Intersection frequency and phase difference |
| $\boldsymbol{G}_{z1}$ | $P_{ref}$ = 0 p.u., $Q_{ref}$ = −1 p.u. | 226 Hz / 74.6° |
| | $P_{ref}$ = −0.95 p.u., $Q_{ref}$ = −0.31 p.u. | 225 Hz / 75.2° |
| | $P_{ref}$ = −0.95 p.u., $Q_{ref}$ = 0.31 p.u. | 224 Hz / 76.1° |
| $\boldsymbol{G}_{z2}$ | $P_{ref}$ = 0 p.u., $Q_{ref}$ = −1 p.u. | 216 Hz/ 69.5° |
| | $P_{ref}$ = −0.95 p.u., $Q_{ref}$ = −0.31 p.u. | 203 Hz / 75.5° |
| | $P_{ref}$ = −0.95 p.u., $Q_{ref}$ = 0.31 p.u. | 197 Hz / 78.2° |
| Robustness analysis against rotor frequency deviations | | |
| Method | Rotor frequency | Intersection frequency and phase difference |
| $\boldsymbol{G}_{z1}$ | 60 Hz | 230 Hz / 74.8° |
| | 40 Hz | 216 Hz / 78.8° |
| $\boldsymbol{G}_{z2}$ | 60 Hz | 201 Hz / 77.6° |
| | 40 Hz | 183 Hz / 82.7° |
| Robustness analysis against leakage inductance deviations | | |
| Method | DFIG leakage inductance deviation | Intersection frequency and phase difference |
| $\boldsymbol{G}_{z1}$ | +20% | 227 Hz / 71.9° |
| | -20% | 226 Hz / 79.0° |
| $\boldsymbol{G}_{z2}$ | +20% | 197 Hz / 75.9° |
| | -20% | 196 Hz / 80.7° |

## V. SIMULATION VERIFICATION

To further verify the effectiveness of the proposed impedance reshaping methods, simulations based on MATLAB/Simulink are carried out. The system parameters are the same as theoretical analysis, which are listed in Table I. Fig. 9(a) shows the simulated waveforms of three-phase stator voltage, stator current, rotor current, as well as the active and reactive power of the DFIG system. In the beginning, $K_{pPLL}$ = 1 p.u., and the system is stable. At $t$ = 2 s, $K_{pPLL}$ increases to 2.57 p.u. and the system loses its stability and an oscillation occurs.

Fast Fourier transformation (FFT) was performed for stator voltage during 2.5 s ~ 2.9 s, the corresponding FFT results are shown in Fig. 9(b). From there, it can be seen that the oscillation frequencies are 114 Hz and 14 Hz, which is consistent with the theoretical analysis results in Fig. 6. Fortunately, when the reshaping method $\boldsymbol{G}_{z1}$ is added at $t$ = 3 s, from Fig. 9(a), it can be seen that the oscillation is significantly suppressed and the system becomes stable. Then switching $\boldsymbol{G}_{z1}$ to $\boldsymbol{G}_{z2}$ at $t$ = 3.5 s, the system still keeps stable, implying that $\boldsymbol{G}_{z2}$ can also suppress the oscillation. The simulation results greatly demonstrate the correctness of the theoretical analysis and the effectiveness of the proposed methods.

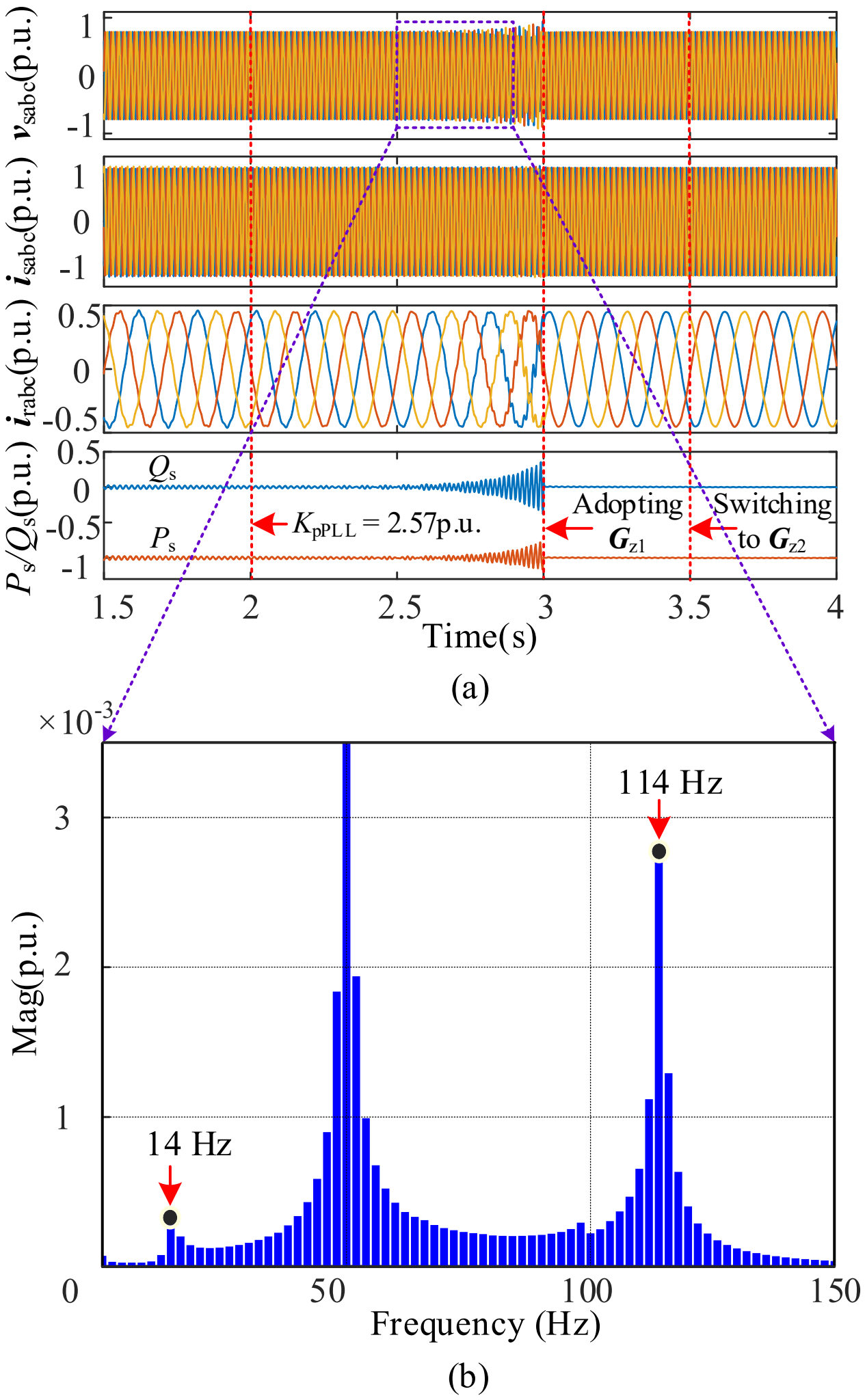

Fig. 9. Simulation waveforms of applying the reshaping methods and FFT analysis results. (a) Simulation waveforms of the DFIG system. (b) FFT analysis results of the stator voltage waveform during 2.5 s ~ 2.9 s.

The robustness of the reshaping methods against the deviations of SCR, power requirements, and rotor frequency are all tested by simulations. The results are shown in Fig. 10~Fig. 12, respectively. The operating conditions of the system before $t$ = 3 s are the same as that in Fig. 9(a). In Fig. 10, $\boldsymbol{G}_{z2}$ is introduced first at $t$ = 3 s, and the oscillation is suppressed and the system operates stably, indicating that $\boldsymbol{G}_{z2}$ can also be used to remove the oscillation caused by PLL. Then SCR jumps from 2 to 10 at $t = 3.5$ s, and the system is still stabilized, which is consistent with the theoretical analysis in Fig. 8. Finally, switching $\boldsymbol{G}_{z2}$ to $\boldsymbol{G}_{z1}$ at $t = 4$ s, the system still keeps stable. The above simulation results prove that both reshaping methods can reduce the influence of PLL and improve the stability margin of the system even when SCR varies.

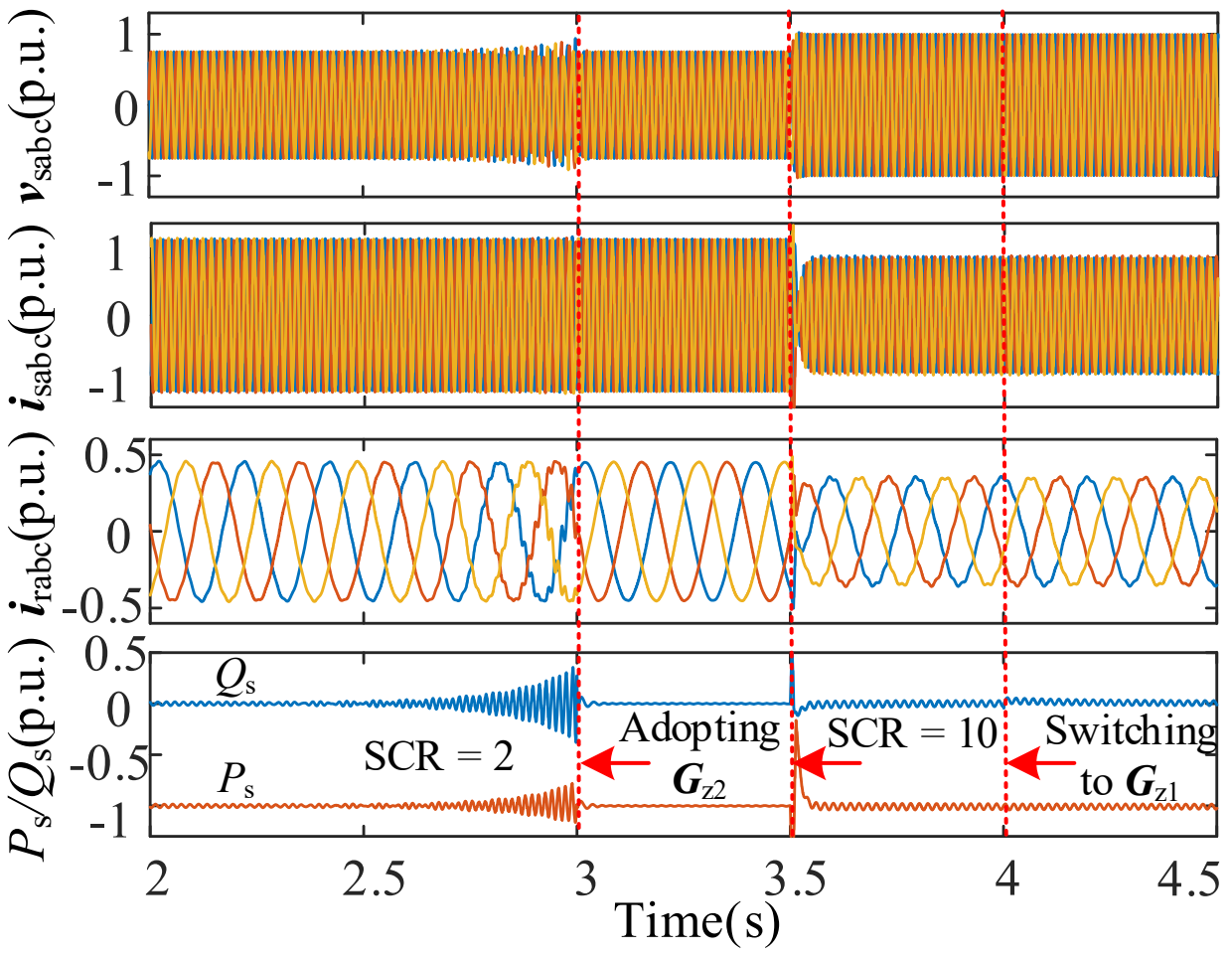

Fig. 10. Simulation results of applying the reshaping methods when SCR varies.

In Fig. 11, $P_{ref}$ is changed from −1 p.u. to 0 p.u. and $Q_{ref}$ is changed from 0 p.u. to −1 p.u. at $t = 3.5$ s. It can be seen that $\boldsymbol{G}_{z1}$ can still effectively suppress the oscillation even if the active and reactive power references are reversed. Later, switching the reshaping method $\boldsymbol{G}_{z1}$ to $\boldsymbol{G}_{z2}$ at $t = 4.5$ s, it can be observed that the system keeps stable either. The simulation results agree with the robust analysis against power requirement deviations in Section IV, indicating both methods can still reduce the influence of PLL and improve the stability margin of the DFIG system when power references vary.

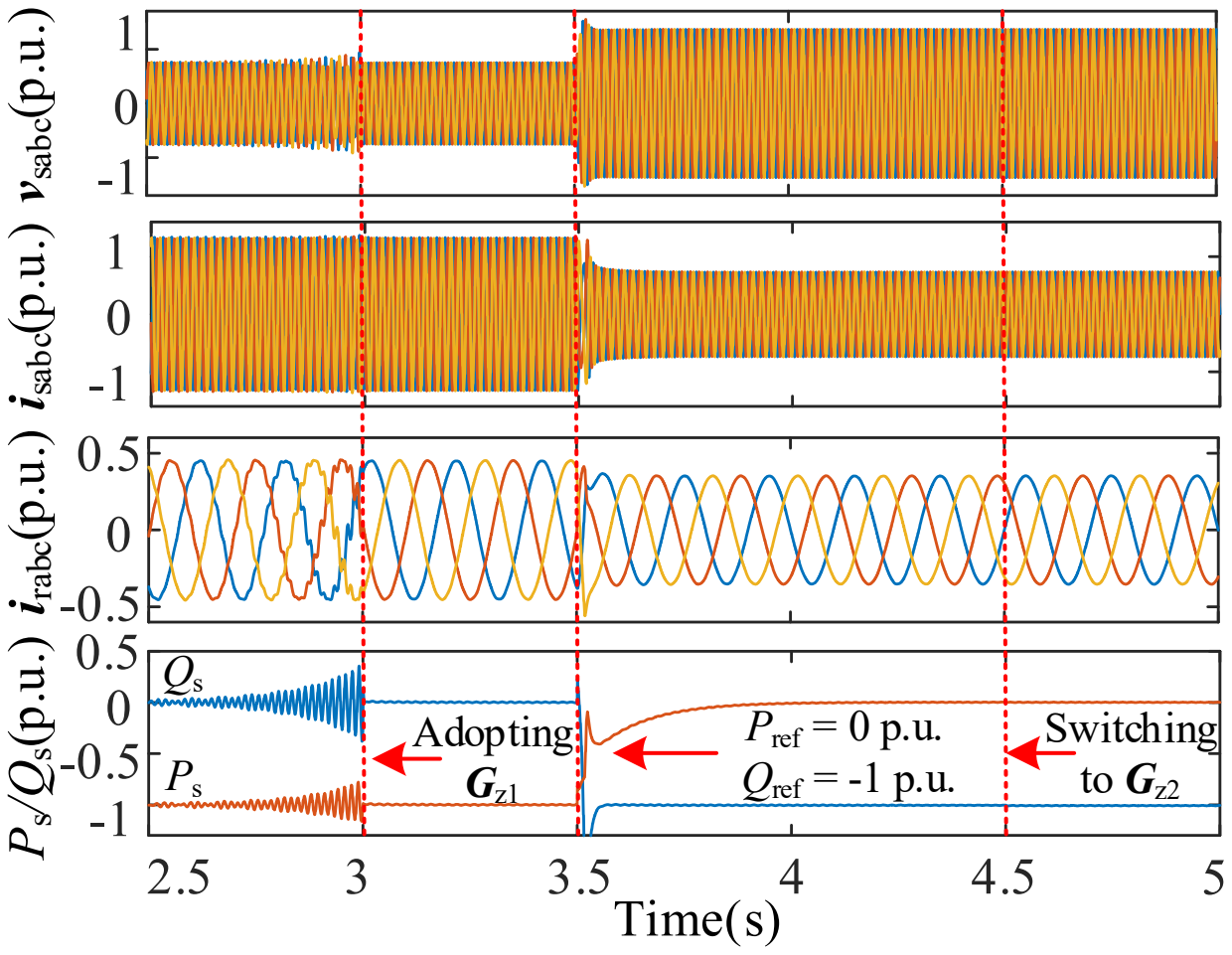

Fig. 11. Simulation results of applying the reshaping methods when DFIG power references vary.

In Fig. 12, at $t = 3.5$ s, the rotor frequency changes from 55 Hz to 40 Hz, and the DFIG system operates from super-synchronous speed to subsynchronous speed. According to Fig. 12, the system remains stable after the rotor frequency varies, and the stator output power is finally adjusted to the expected values. Then, changing the reshaping method from $\boldsymbol{G}_{z1}$ to $\boldsymbol{G}_{z2}$ at $t = 4.5$ s, the system is also stabilized. The result verifies the robustness analysis against rotor frequency deviations, showing that both methods still have good performance during rotor frequency variations. These simulation results further validate the effectiveness and enhancement of the proposed methods for system stability.

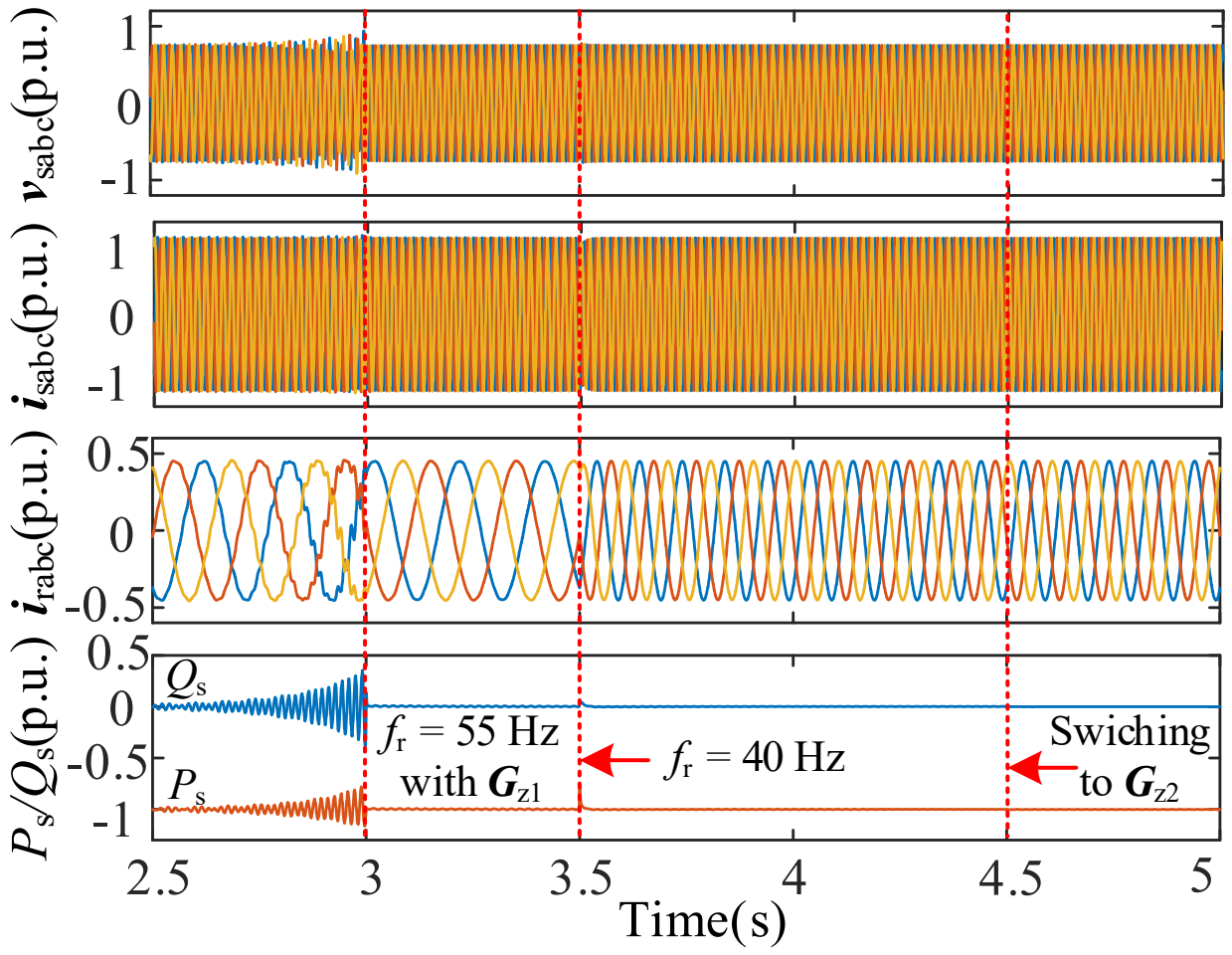

Fig. 12. Simulation results of applying the reshaping methods when rotor frequency varies.

## VI. Conclusion

In this paper, the impact of PLL is analyzed through the MIMO impedance model of DFIG system. It is found that the frequency coupling phenomena is mainly caused by the transfer function matrix linked to rotor current dynamic. Thus, in order to eliminate the influence of rotor current dynamic caused by PLL and improve the system stability, a second-order impedance reshaping method based on compensating rotor current dynamic is proposed, which can eliminate the frequency coupling effect under various operation conditions. Moreover, the negative resistance region caused by PLL was removed, and thus the system stability was improved a lot. Furthermore, a first-order compensation method is proposed based on simplifying the second-order reshaping method, which can also effectively suppress the oscillations, and can be easily implemented. Next, robust analysis is given to illustrate that both the proposed methods can improve the stability margin of the system when the system operating conditions and parameters vary. Finally, simulation validations are also carried out in MATLAB/Simulink, and the results verified the effectiveness of the theoretical analysis and the improvement for system stability of the proposed methods.

## References

[1] H. Polinder, J. A. Ferreira, B. B. Jensen, A. B. Abrahamsen, K. Atallah and R. A. McMahon, “Trends in Wind Turbine Generator Systems,” *IEEE J. Emerg. Sel. Topics Power Electron*., vol. 1, no. 3, pp. 174-185, Sept. 2013.

[2] R. Cardenas, R. Pena, S. Alepuz, and G. Asher, “Overview of control systems for the operation of DFIGs in wind energy applications,” *IEEE Trans. Ind. Electron*., vol. 60, no. 7, pp. 2776-2798, Jul, 2013.

[3] Y. Song and F. Blaabjerg, “Analysis of middle frequency resonance in DFIG system considering phase locked loop,” *IEEE Trans. Power Electron*., vol. 33, no. 1, pp. 343–356, Jan. 2018.

[4] C. Zhang, X. Cai, M. Molinas, and A. Rygg, “On the impedance modeling and equivalence of AC/DC side stability analysis of a grid-tied Type IV wind turbine system,” *IEEE Trans. Energy Convers*., vol. 34, no. 2, pp. 1000–1009, Jun. 2019.

[5] H. Zhang, X. Wang, L. Harnefors, H. Gong, J. P. Hasler, et al, “SISO transfer functions for stability analysis of grid-connected voltage-source converters,” *IEEE Trans. Ind. Appl*., vol. 55, no. 3, pp. 2931-2941, May/Jun. 2019.

[6] Y. Xu, H. Nian, T. Wang, L. Chen, and T. Zheng, “Frequency coupling characteristic modeling and stability analysis of doubly fed induction generator,” *IEEE Trans. Energy Convers*., vol. 33, no. 3, pp. 1475–1486, Sep. 2018.

[7] Y. Sun, X. Zhang, M. Han, F. Xiao, J. Yu and H. Zhang, “General Impedance Model of DFIG for Wide-Range-Frequency Oscillation Studies,” *IEEE 9th International Power Electronics and Motion Control Conference*, Nanjing, China, Nov. 2020, pp. 2882-2886.

[8] B. Wen, D. Boroyevich, R. Burgos, P. Mattavelli and Z. Shen, “Analysis of D-Q Small-Signal Impedance of Grid-Tied Inverters,” *IEEE Trans. Power Electron.*, vol. 31, no. 1, pp. 675-687, Jan. 2016.

[9] B. Hu, H. Nian, M. Li, Y. Liao, J. Yang and H. Tong, “Impedance Characteristic Analysis and Stability Improvement Method for DFIG System Within PLL Bandwidth Based on Different Reference Frames,” *IEEE Trans. Ind. Electron.*, vol. 70, no. 1, pp. 532-543, Jan. 2023.

[10] L. Fan and Z. Miao, "Mitigating SSR Using DFIG-Based Wind Generation," *IEEE Trans. Sustainable Energy*, vol. 3, no. 3, pp. 349-358, July 2012.

[11] A. E. Leon, "Integration of DFIG-Based Wind Farms into Series-Compensated Transmission Systems," *IEEE Trans. Sustainable Energy*, vol. 7, no. 2, pp. 451-460, April 2016.

[12] I. Vieto and J. Sun, “Refined small-signal sequence impedance models of type-III wind turbines,” in *Proc. IEEE Energy Convers. Expo.*, Sep. 2018, pp. 2242–2249.

[13] X. Wang, Y . W. Li, F. Blaabjerg, and P . C. Loh, “Virtual-impedance-based control for voltage-source and current-source converters,” *IEEE Trans. Power Electron.*, vol. 30, no. 12, pp. 7019–7037, Dec. 2015.

[14] S. Xu, X. Wu, W. Gu, L. Fan, Y. Lu and Z. Zou, "Mitigating Subsynchronous Oscillation Using Adaptive Virtual Impedance Controller in DFIG Wind Farms," *IEEE Sustainable Power and Energy Conference* (*iSPEC*), Nanjing, China, 2021, pp. 1801-1807.

[15] B. Hu, H. Nian, J. Yang, M. Li and Y. Xu, "High-Frequency Resonance Analysis and Reshaping Control Strategy of DFIG System Based on DPC," *IEEE Trans. Power Electron.*, vol. 36, no. 7, pp. 7810-7819, July 2021.

[16] H. Tong, H. Nian, B. Hu, M. Li, H. Zhang and Q. Liu, "High-Frequency Resonance Analysis Between DFIG Based Wind Farm with Direct Power Control and VSC-HVDC," *24th International Conference on Electrical Machines and Systems* (*ICEMS*), Gyeongju, Korea, Dec. 2021, pp. 2207-2212.

[17] D. Yang, X. Wang, F. Liu, K. Xin, Y. Liu and F. Blaabjerg, “Symmetrical PLL for SISO Impedance Modeling and Enhanced Stability in Weak Grids,” *IEEE Trans. Power Electron.*, vol. 35, no. 2, pp. 1473-1483, Feb. 2020.

[18] C. Wu, B. Hu, P. Cheng, H. Nian and F. Blaabjerg, “Eliminating Frequency Coupling of DFIG System Using a Complex Vector PLL,” *IEEE Applied Power Electronics Conference and Exposition*, New Orleans, LA, USA, Mar. 2020, pp. 3262-3266.

[19] H. Nian, B. Hu, Y. Xu, C. Wu, L. Chen and F. Blaabjerg, “Analysis and Reshaping on Impedance Characteristic of DFIG System Based on Symmetrical PLL,” *IEEE Trans. Power Electron.*, vol. 35, no. 11, pp. 11720-11730, Nov. 2020.

[20] Y. Zhang, C. Klabunde and M. Wolter, “Frequency-Coupled Impedance Modeling and Resonance Analysis of DFIG-Based Offshore Wind Farm With HVDC Connection,” *IEEE Access*, vol. 8, pp. 147880-147894, Aug. 2020.

[21] X. Zhang, Y. Zhang, R. Fang and D. Xu, “Impedance Modeling and SSR Analysis of DFIG Using Complex Vector Theory,” *IEEE Access*, vol. 7, pp. 155860-155870, Oct. 2019.

[22] X. Wang, L. Harnefors and F. Blaabjerg, “Unified Impedance Model of Grid-Connected Voltage-Source Converters,” *IEEE Trans. Power Electron.*, vol. 33, no. 2, pp. 1775-1787, Feb. 2018.

[23] B. Hu, H. Nian, M. Li, Y. Xu, Y. Liao and J. Yang, “Impedance-Based Analysis and Stability Improvement of DFIG System Within PLL Bandwidth,” *IEEE Trans. Ind. Electron*., vol. 69, no. 6, pp. 5803-5814, June 2022.

[24] B. Liang, J. He, Y. W. Li, P. Guo and C. Wang, “Aggregated-Impedance-Based Stability Analysis for a Parallel-Converter System Considering the Coupling Effect of Voltage Feedforward Control and Reactive Power Injection,” *IEEE Trans. Power Electron.*, vol. 36, no. 5, pp. 5954-5970, May 2021.

[25] A. J. Agbemuko, J. L. Domínguez-García, O. Gomis-Bellmunt and L. Harnefors, “Passivity-Based Analysis and Performance Enhancement of a Vector Controlled VSC Connected to a Weak AC Grid,” *IEEE Trans. Power Del*., vol. 36, no. 1, pp. 156-167, Feb. 2021.

[26] L. Huang, C. Wu, D. Zhou and F. Blaabjerg, "Comparison of DC-link Voltage Control Schemes on Grid-side and Machine-side for Type-4 Wind Generation System Under Weak Grid," *IECON* 2021 – 47*th Annual Conference of the IEEE Industrial Electronics Society*, Toronto, Canada, Oct. 2021, pp. 1-6.

[27] L. Huang, C. Wu, D. Zhou and F. Blaabjerg, "A Simplified SISO Small-Signal Model for Analyzing Instability Mechanism of Grid-Forming Inverter under Stronger Grid," 2021 *IEEE* 22*nd Workshop on Control and Modelling of Power Electronics*, Cartagena, Colombia, Nov. 2021, pp. 1-6.

[28] C. Zhang, X. Cai, A. Rygg, and M. Molinas, “Sequence domain SISO equivalent models of a grid-tied voltage source converter system for small-signal stability analysis,” *IEEE Trans. Energy Convers.*, vol. 33, no. 2,pp. 741–749, Jun. 2018